\documentclass[manuscript]{acmart}
\AtBeginDocument{%
  }

\setcopyright{acmlicensed}
\copyrightyear{2026}
\acmYear{2026}
\acmDOI{XXXXXXX.XXXXXXX}
\acmConference[CHI'26]{The ACM Association of Computing Machinery}{April 13--17, 2026}{Barcelona, Spain}
\acmISBN{978-1-4503-XXXX-X/2018/06}
\acmISBN{978-1-4503-XXXX-X/2018/06}

\begin{document}

\title{Human-AI Interaction Traces as Blackout Poetry: Reframing AI-Supported Writing as Found-Text Creativity}

\author{Syemin Park}
\email{syeminpark@kaist.ac.kr}
\affiliation{%
  \institution{Department of Industrial Design, KAIST}
  \city{Daejeon}
  \country{Republic of Korea}
}

\author{Soobin Park}
\email{soobinpark@kaist.ac.kr}
\affiliation{%
  \institution{Department of Industrial Design, KAIST}
  \city{Daejeon}
  \country{Republic of Korea}
}

\author{Youn-kyung Lim}
\email{younlim@kaist.ac.kr}
\affiliation{%
  \institution{Department of Industrial Design, KAIST}
  \city{Daejeon}
  \country{Republic of Korea}
}

\renewcommand{\shortauthors}{Park et al.}

\begin{abstract}
LLMs offer new creative possibilities for writers but also raise concerns about authenticity and reader trust, particularly when AI involvement is disclosed. Prior research has largely framed this as an issue of transparency and provenance, emphasizing the disclosure of human–AI interaction traces that account for how much the AI wrote and what the human did. Yet such audit-oriented disclosures may risk reducing creative collaboration to quantification and surveillance. In this position paper, we argue for a different lens by exploring how human–AI interaction traces might instead function as expressive artifacts that foreground the meaning-making inherent in human–AI collaboration. Drawing inspiration from blackout poetry, we frame AI-generated text as found material through which writers’ acts of curation and reinterpretation become inscribed atop the AI’s original output. In this way, we suggest that designing interaction traces as aesthetic artifacts may help readers better appreciate and trust writers’ creative contributions in AI-assisted writing.
\end{abstract}


\begin{teaserfigure}
  \centering
  \includegraphics[width=0.5\textwidth]{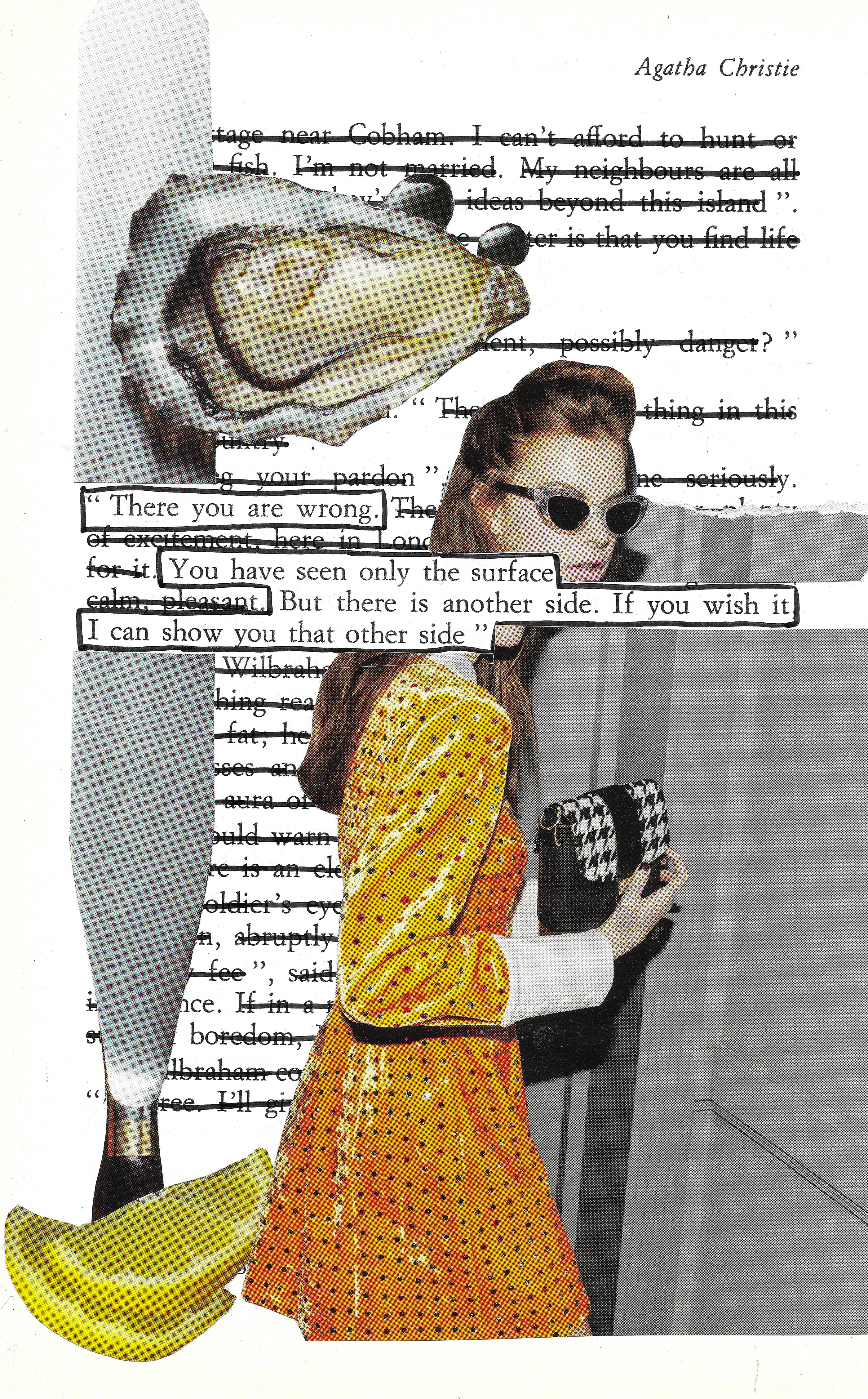}
  \caption{A blackout poetry example \cite{macchini_you_have_only_seen_the_surface_2026} © Giulia Macchini.}
  \label{fig:teaser}
\end{teaserfigure}


\maketitle

\section{Position Statement}

AI-supported writing has been rapidly transforming how creative work is produced. While these tools can offer writers meaningful benefits, such as facilitating ideation \cite{10.1145/3715336.3735832, 10.1145/3511599} and providing resources for interpretation \cite{10.1145/3796234,10.1145/3635636.3656187}, they also introduce new tensions. In particular, recent studies indicate that AI disclosure often elicits negative perceptions among readers \cite{li2024does}, and readers with more negative attitudes toward AI exhibit a greater decline in trust when AI involvement is disclosed \cite{schilke2025transparency}.
Given that writing can be seen as “\textit{a social act that conveys an author's intention, effort, and emotion to readers}” \cite{nakano2025understanding}, writers now face new challenges in communicating AI involvement in ways that maintain reader trust and preserve perceptions of authentic expression \cite{10.1145/3544548.3580782}.

In response to these challenges, HCI researchers have begun exploring systems that make AI involvement in writing more transparent by capturing traces of the AI-assisted writing process. For instance, HaLLMark \cite{10.1145/3613904.3641895} is a visualization system that tracks AI prompts over time and highlights the associated AI-written or AI-influenced text within a writing interface. In addition, DraftMark \cite{siddiqui2025draftmarks} utilizes visualization methods inspired by physical metaphors like eraser crumbs to indicate sections in the final text that received writers' revision effort. While these works offer promising directions, writers have also shown concerns about making AI use completely transparent, as they fear such disclosure may invite harsh criticism or devalue their creative contribution \cite{10.1145/3613904.3641895, 10.1145/3711020}. Consequently, recent scholarship  \cite{10.1145/3613904.3641895, 10.1145/3711020} has called for greater attention to the perceptual frameworks that readers bring to work created with AI assistance, encouraging more open-minded interpretations of human–AI collaboration.

Thus, instead of focusing on audit-oriented transparency tools that may invite undue scrutiny, we explore designs of activity traces that foreground writers’ creativity inherent in human–AI collaboration, helping mitigate negative biases in readers' perceptions of AI-assisted writing. To do this, we take inspiration from blackout poetry. Blackout poetry, as a form of found art \cite{
stribling1970art,dayton1999art,tankersley2009found}, has the “goal of using pre-existing or “found” text to create something new. Portions of the original text are kept, while other parts are drawn over or crossed out.” \cite{sccld2023blackout}. As seen in \autoref{fig:teaser}, blackout poetry often also incorporates visualization practices to construct new narratives and interpretations from pre-existing materials.

Crucially, in blackout poetry, the creative contribution lies not in writing new sentences from scratch but in the transformative act of curation: deciding which words to keep. This form of subtractive authorship resonates with how writers themselves conceptualize contribution in AI-assisted writing. Specifically, prior research \cite{10.1145/3711020} finds that writers often view AI-supported writing as authentic as long as they are actively deciding (e.g., selecting, adopting, removing) what ultimately enters the final text (content gatekeeping), rather than how much text they actually wrote.

Thus, we argue that AI-generated text can be framed as a form of found text, and the human writer’s role can be presented to readers as a creative sculptor of this material. The body of initially generated AI text would then become trace-bearing substrates that inscribe acts of selection (e.g., visualizing blackouts) and appropriation (e.g., abstract drawings that foreground the writer’s interpretive framing of the retained text). In this way, these trace-based representations can function as aesthetic media through which readers may better perceive and appreciate writers' intention, taste, and meaning-making when working with AI-generated materials.

Finally, we note that certain prerequisites may be necessary to create blackout poetry-inspired traces. This visualization approach depends on AI-generated text functioning as a body of found material that writers can subsequently curate, reshape, and appropriate. When AI output is produced as polished, “perfect” prose, it may leave limited room for meaningful intervention or creative transformation. Recent work has therefore proposed generating AI output in intermediary formats \cite{10.1145/3796234,10.1145/3635636.3656187}, rather than as finalized text, in order to support writers’ reflection, interpretation, and adaptation. Such generative substrates may provide especially fertile ground for blackout-style traces, enabling writers to visibly inscribe their creative and interpretive agency within the evolving text.

To summarize, drawing inspiration from blackout poetry, we propose a reframing of AI-assisted writing in which AI-generated text becomes found material transformed through writers’ subtractive authorship. We hope this work inspires further research that enables human–AI interaction traces to serve as an expressive channel through which writers can communicate their intention, taste, and interpretive agency as part of their creative practice.

\begin{acks}
We are grateful to Forking Room (Min-hyung Kang, Soo Yon Song, and Binna Choi) for inspiring this work. This work was supported by the National Research Foundation of Korea (NRF) grant funded by the Korea government (MSIT) (No. RS-2021-NR059056).

\end{acks}

\bibliographystyle{ACM-Reference-Format}
\bibliography{reference}

\end{document}